\documentclass[]{spie}  

\usepackage{amsmath,amsfonts,amssymb}
\usepackage{graphicx}
\usepackage[colorlinks=true, allcolors=blue]{hyperref}
\usepackage{xfrac} 

\title{A-STEP: The AstroPix Sounding Rocket Technology Demonstration Payload}

\author[a,b]{Daniel P. Violette}
\author[a,b]{Amanda Steinhebel}
\author[c]{Abhradeep Roy}
\author[a]{Ryan Boggs}
\author[a]{Regina Caputo}
\author[a]{David Durachka}
\author[c]{Yasushi Fukazawa}
\author[c]{Masaki Hashizume}
\author[a,d]{Scott Hesh}
\author[e]{Manoj Jadhav}
\author[a]{Carolyn Kierans}
\author[a,f]{Kavic Kumar}
\author[g]{Shin Kushima}
\author[h]{Richard Leys}
\author[e]{Jessica Metcalfe}
\author[a,f]{Zachary Metzler}
\author[c]{Norito Nakano}
\author[h]{Ivan Peric}
\author[a]{Jeremy Perkins}
\author[a,d]{Lindsey Seo}
\author[i]{K.-W. Taylor Shin}
\author[h]{Nicolas Striebig}
\author[c]{Yusuke Suda}
\author[g]{Hiroyasu Tajima}

\affil[a]{NASA Goddard Space Flight Center, Greenbelt, MD, USA}
\affil[b]{NASA Postdoctoral Program Fellow (ORAU)}
\affil[c]{Physics Program, Graduate School of Advanced Science and Engineering, Hiroshima University, Hiroshima, Japan}
\affil[d]{NASA Wallops Flight Facility, Wallops Island, VA, USA}
\affil[e]{Argonne National Laboratory, Lemont, IL, USA}
\affil[f]{Department of Physics, University of Maryland, College Park, MD, USA}
\affil[g]{Institute for Space–Earth Environmental Research, Nagoya University, Aichi, Japan}
\affil[h]{ASIC and Detector Laboratory, Karlsruhe Institute of Technology, Karlsruhe, Germany}
\affil[i]{Institute for Particle Physics, University of California, Santa Cruz, CA, USA}

\authorinfo{Further author information: (Send correspondence to Daniel Paul Violette)\\E-mail: daniel.p.violette@nasa.gov}

\begin{document} 
\maketitle

\begin{abstract}
A next-generation medium-energy (100\,keV to 100\,MeV) gamma-ray observatory will greatly enhance the identification and characterization of multimessenger sources in the coming decade. Coupling gamma-ray spectroscopy, imaging, and polarization to neutrino and gravitational wave detections will develop our understanding of various astrophysical phenomena including compact object mergers, supernovae remnants, active galactic nuclei and gamma-ray bursts. An observatory operating in the MeV energy regime requires technologies that are capable of measuring Compton scattered photons and photons interacting via pair production. AstroPix is a monolithic high voltage CMOS active pixel sensor which enables future gamma-ray telescopes in this energy range. AstroPix's design is iterating towards low-power ($\sim$1.5\,mW/cm$^{2}$), high spatial (500\,$\mu$m pixel pitch) and spectral ($<$5\,keV at 122\,keV) tracking of photon and charged particle interactions. Stacking planar arrays of AstroPix sensors in three dimensions creates an instrument capable of reconstructing the trajectories and energies of incident gamma rays over large fields of view. A prototype multi-layered AstroPix instrument, called the AstroPix Sounding rocket Technology dEmonstration Payload (A-STEP), will test three layers of AstroPix “quad chips” in a suborbital rocket flight. These quad chips (2$\times$2 joined AstroPix sensors) form the 4$\times$4\,cm$^{2}$ building block of future large area AstroPix instruments, such as ComPair-2 and AMEGO-X. This payload will be the first demonstration of AstroPix detectors operated in a space environment and will demonstrate the technology’s readiness for future astrophysical and nuclear physics applications. In this work, we overview the design and state of development of the A-STEP payload. 
\end{abstract}


\section{INTRODUCTION}
\label{sec:introduction}  

Multi-messenger observations are a powerful method to study the most extreme astrophysical phenomena (compact mergers, accelerators, and explosions) and are an important driver for the astronomy community, as outlined in the 2020 Astronomy and Astrophysics Decadal Survey\cite{ASTRO2020}. Gamma-ray observatories provide the crucial electromagnetic counterpart for many astrophysical sources that generate neutrinos, cosmic rays, and gravitational waves. Despite this need, there remains a lack of next-generation wide-field, high-resolution, high-sensitivity instruments capable of imaging gamma-ray sources, especially in the medium energy gamma-ray (100\,keV to 100\,MeV) band.

The medium energy gamma-ray regime requires an observatory design capable of capturing Compton scattering and pair-conversion photon interactions. To accomplish this, large area multi-layered “tracker” telescopes are required, capable of accurately reconstructing particle trajectories by resolving the three-dimensional position and energy information left by particle and photon tracks passing through the instrument. To date, these tracker systems have largely been designed with silicon strip detectors (SSDs). Each one-dimensional charge-collecting strip is bonded to an individual readout channel in an ASIC for signal amplification and digitization. To achieve two-dimensional position knowledge, these strips must be layered orthogonally in adjacent detector layers. Additionally, the long silicon strips inherently have high capacitance noise. A tracker telescope design that improves upon both position and energy resolution over the capabilities of legacy systems will be a requirement for any future gamma-ray observatory.

\subsection{AstroPix: HV-CMOS MAPS}
\label{subsec:astropix}

High Voltage CMOS Monolithic Active Pixel Sensors (HV-CMOS MAPS)\cite{Peric18} are a promising technology for future tracker telescopes. Originally designed for collider physics experiments, HV-CMOS MAPS pixels inherently provide the two-dimensional position knowledge lacking from SSDs. Additionally, the CMOS design incorporates signal amplification and digitization within each pixel, thus eliminating the need for an external readout ASIC and reducing capacitance noise and power consumption. These elements lend themselves to tracker telescope designs that require large multi-layered arrays of detectors. 

AstroPix is a HV-CMOS MAPS based off of earlier ATLASPix detectors designed for the ATLAS experiment at the European Organization for Nuclear Research (CERN)\cite{Brewer2021}. Signal amplification and digitization (via a precise time-over-threshold pulse measurement) occurs on-pixel and is matched with a timestamp at the hit buffer during readout. Earlier versions of AstroPix (through {\tt AstroPixv3}, flying on A-STEP) have row and column hit buffers that are paired to capture a full event, while newer AstroPix versions have individual hit buffers for each pixel. Each 2$\times$2\,cm$^{2}$ AstroPix detector has 35$\times$35 (500\,$\mu$m pitch) pixels operating at low power ($\sim$1.5\,mW/cm$^{2}$ combined digital and analog). AstroPix commanding and data collection is handled over a quasi-SPI interface. When a photon event occurs, a separate digital interrupt line shared by each AstroPix chip on the quad chip signals the frontend firmware to start the SPI clock. While the clock is running, digital data is passed from chip to chip along each digital periphery in the quad chip. This readout is scaleable beyond a single quad chip, enabling tight tiling of large numbers of detectors with data and commands daisy-chained between sensors to and from the array edge. The AstroPix development program has benefited from rapid, incremental detector design. More information on {\tt AstroPixv3} can be found in Steinhebel et al. 2022\cite{Steinhebel2022}, and information on {\tt AstroPixv4} can be found in Striebig et al. 2024\cite{Striebig2024}. Future versions of AstroPix are expected to fully deplete the 500\,$\mu$m thick pixels and enable a 25-700\,keV dynamic energy range.

\subsection{Applications of AstroPix Detectors}

One potential application of the AstroPix detector technology is the All-Sky Medium Energy Gamma ray Observatory eXplorer (AMEGO-X) concept\cite{Caputo2022}. AMEGO-X is a Medium-Class Explorer (MIDEX) mission concept utilizing three instruments (AstroPix tracker telescope, cesium iodide (CsI) calorimeter, and an active anti-coincidence particle detector) to identify and reconstruct medium-energy gamma rays from 25\,keV to 1\,GeV. AMEGO-X would achieve an order magnitude improvement in medium-energy gamma-ray sensitivity, particularly around the 1 to 10\,MeV band. AMEGO-X’s tracker telescope would be composed of four individual towers, each with 40 layers of 380 AstroPix detectors. The integration and operation of such a large system of detectors is made possible by AstroPix's integrated digitization, chain-able readout and low-power MAPS design. 

AstroPix detectors development has also yielded benefits for nuclear physics collider experiments. AstroPix has been selected for use in Brookhaven National Laboratory's Electron-Ion Collider (EIC)\cite{Apadula2022}, specifically for the electron-Proton Ion Collider (ePIC) detector, targeted to begin operation in the 2030s. AstroPix will be a critical component of ePIC's Imaging Barrel Calorimeter.

\begin{figure}[ht]
    \centering
    \includegraphics[width=0.5\textwidth]{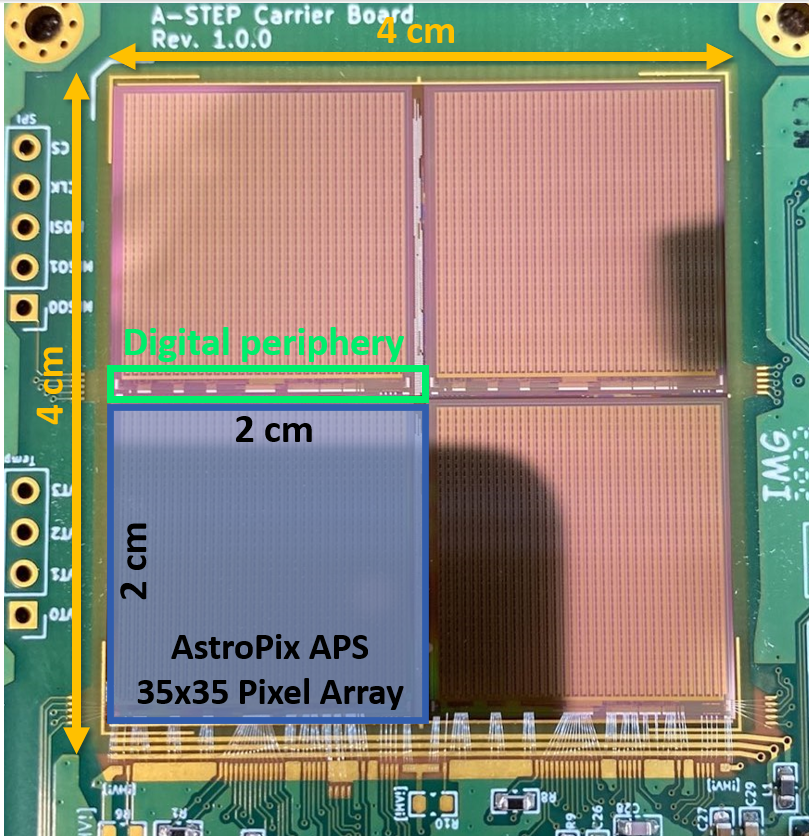}
    \caption{AstroPix quad chip: 2$\times$2 AstroPix Active Pixel Sensors (APS) mounted to an AstroPix Carrier Board.}
    \label{fig:fig1}
\end{figure}

The AstroPix Sounding rocket Technology dEmonstration Payload (A-STEP) will be the first \textit{in-situ} test of {\tt AstroPixv3} detectors in a space environment and demonstrate the viability of AstroPix sensor technology for space-based gamma-ray observatories. This technology demonstration payload is manifested on the SubTEC-10 sounding rocket mission managed out of Wallops Flight Facility and Goddard Space Flight Center. The design of A-STEP resembles a miniature tracker telescope, incorporating three AstroPix detector layers, each containing a ``quad chip'' (2$\times$2 undiced AstroPix sensors). Figure \ref{fig:fig1} depicts a mounted AstroPix quad chip and labels the 35$\times$35 pixel grid and a chip's digital periphery. In this paper, we will outline the full A-STEP missions, including mission objectives, payload design, integration and testing plans, simulation studies, and current schedule.

\section{A-STEP Overview}
\label{sec:overview}

\subsection{A-STEP Mission Objectives}

The top-level requirement of A-STEP is to raise the technology readiness level (TRL) of the AstroPix HV-CMOS MAPS for future medium-energy gamma-ray observatories. This can be achieved by operating a single AstroPix quad chip in the space environment provided by a short duration sounding rocket flight. By receiving a combination of telemetered AstroPix pixel data and payload housekeeping information, we can confirm the successfully operation of AstroPix detectors in a relevant environment. By tracking thermal conditions, HV bias voltage, and power consumption, we will identify trends in performance over the course of the sounding rocket flight.

A-STEP has two secondary goals that will be important for the development of tracker telescopes utilizing AstroPix technology. A-STEP seeks to reconstruct particle tracks as they pass through multiple layers of AstroPix detectors. To achieve this goal, A-STEP must be capable of returning reliable matching row and column hits with coincident timestamps from each AstroPix quad chip layer. Analysis will be performed post-flight to identify these tracks, enhanced by successful pre-flight pixel-level calibration of the AstroPix quad chips. Utilizing these reconstructed particle tracks, A-STEP will attempt to measure the detected cosmic-ray rates, and match them with effective detector area calculations and the accompanying simulation efforts in Section \ref{sec:simulations}. 

\subsection{A-STEP Payload Design}

\begin{figure}[ht]
    \centering
    \includegraphics[width=0.5\textwidth]{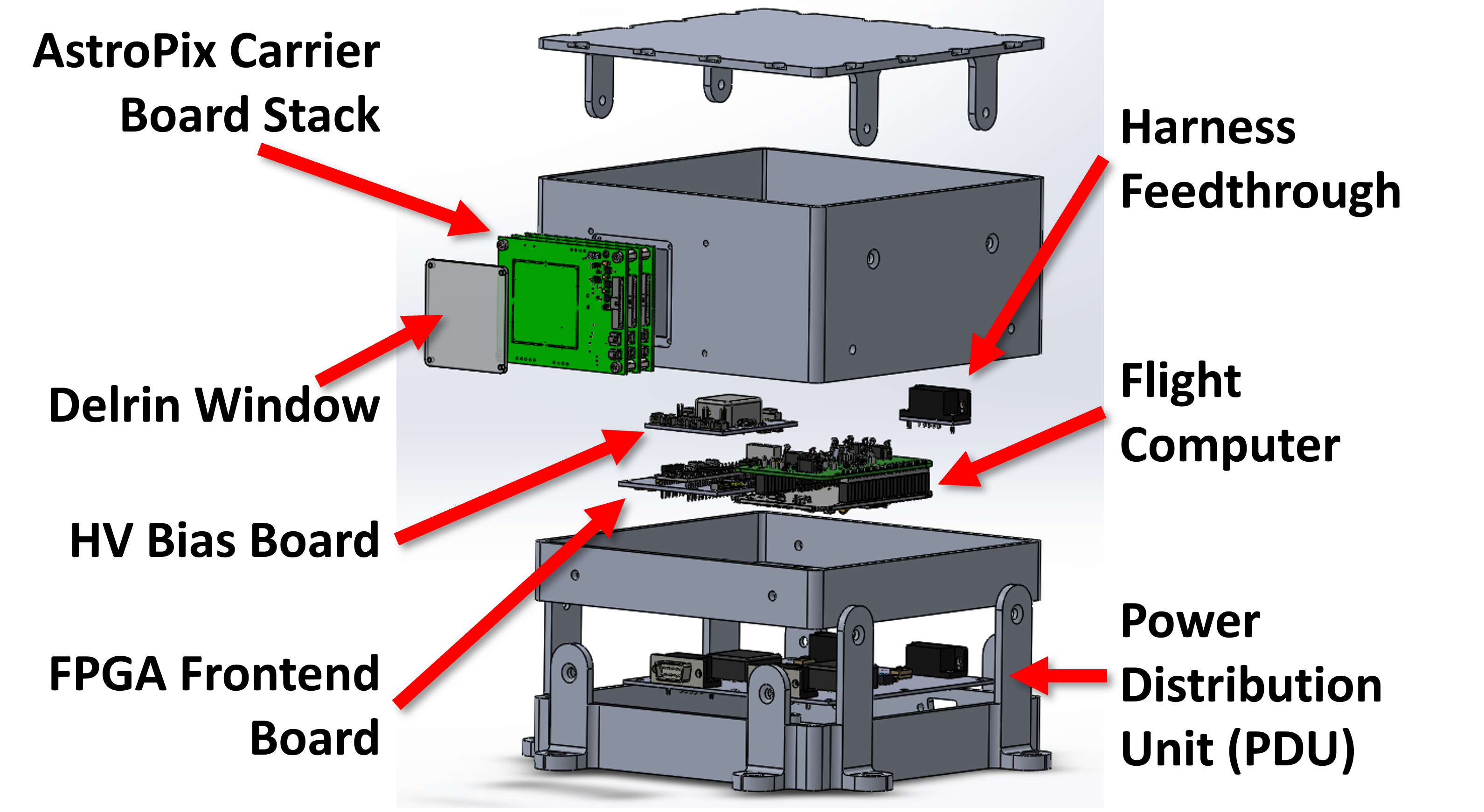}
    \caption{Exploded view of A-STEP payload mechanical enclosure. Mechanical payload design incorporates three tray design to simplify integration and improve heat dissipation.}
    \label{fig:fig2}
\end{figure}

The A-STEP instrument payload has been designed within an aluminum (6061-T6) multi-tray enclosure design that allows for independent mounting and swapping-out of each of the payload's subsystems. Figure \ref{fig:fig2} shows an exploded view of the entire 13.35$\times$20.78$\times$20.78\,cm$^{3}$ payload model, including the payload lid and the three primary payload trays. The top tray incorporates an optically opaque Delrin window and the AstroPix carrier boards. The middle tray contains the BeagleBone Black single-board flight computer, the detector HV biasing board, and the Field Programmable Gate Array (FPGA) frontend board. The bottom tray contains the payload's power dissipation unit (PDU). The payload interfaces with the sounding rocket through an Ethernet via telemetry (EVTM) port for data downlink, and a 28\,V rail input to the A-STEP payload's PDU. The total mass of the payload's mechanical enclosure is less than 4.5\,kg and designed with large margins of safety for both shear and tensile loads on the selected stainless steel bolts and fasteners. The payload will be placed with the AstroPix quad chips and Delrin window facing the $\sfrac{1}{8}$\,in-thick sounding rocket skin to minimize passive shielding that may occur from other sounding rocket deck plates and payloads.

\subsubsection{AstroPix Carrier Board}

\begin{figure}[ht]
    \centering
    \hspace*{2cm}\includegraphics[width=0.8\textwidth]{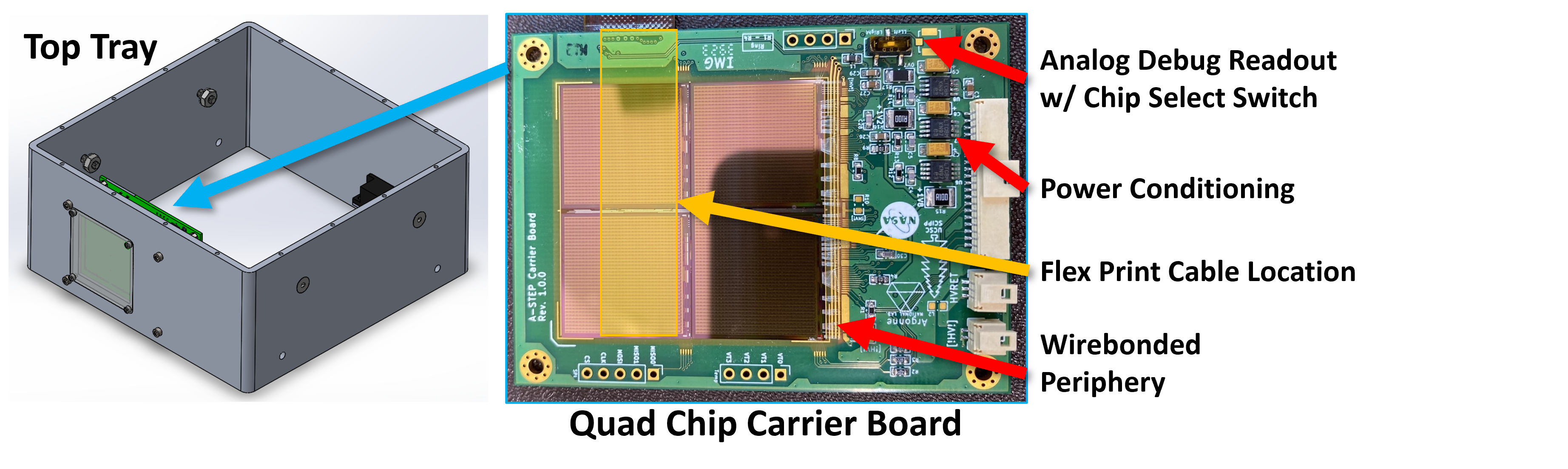}
    \caption{\textit{Left: }Top layer of A-STEP payload, containing a stack of three AstroPix quad chip carrier boards against a Delrin window. \textit{Right: }AstroPix quad chip carrier board.}
    \label{fig:fig3}
\end{figure}

The A-STEP instrument is a stack of three AstroPix quad chip carrier boards mounted on the top payload tray wall behind a visibly opaque Delrin window. The AstroPix carrier boards and top payload tray can be seen in Figure \ref{fig:fig3}. The AstroPix carrier board attachment orientation was chosen to allow the AstroPix detectors to face out through the sounding rocket's skin aluminum skin to maximize charged particle and photon event rates. The AstroPix carrier board printed circuit board (PCB) has a large cut-out over which the AstroPix quad chip is mounted. The edges of the quad chip is epoxied along a 2\,mm overlap with the PCB. This removes excess passive material between the three AstroPix carrier board layers. While a minor improvement for this technology demonstration, proving the mechanical robustness of this design is important for future full-scale tracker telescopes, where any excess passive material can easily accumulate over many layers of detectors. The four AstroPix detectors within the quad chip are wirebonded either directly to the AstroPix carrier board PCB or to a flexprint cable that is epoxied over the top of the chips when the digital peripheries are not adjacent to the PCB (pictured in Figure \ref{fig:fig3}).

On the AstroPix carrier board, the AstroPix sensors are powered with 1.8\,V analog and digital inputs conditioned with low-dropout (LDO) regulators from the board's 4.5\,V input. Each AstroPix chip's configuration can be set either with the board's quasi-SPI data interface, or a back-up shift register interface that additionally allows for reading out loaded configurations. The AstroPix chips' $-$150\,V bias (with an expected $-$40\,nA leakage current per quad chip) is provided through separate latched connectors from the HV supply board, while the remaining power, commanding, and data handling is provided by the FPGA frontend board through a latched 20-pin connector. The AstroPix carrier board also possesses an analog chip readout with a chip select switch for debugging and data quality checks, along with additional mount points for debugging the AstroPix quad chip's quasi-SPI interface or for reading several of the quad-chips internal thermal sensing circuits.

\subsubsection{Frontend Electronics}

\begin{figure}[ht]
    \centering
    \includegraphics[width=0.8\textwidth]{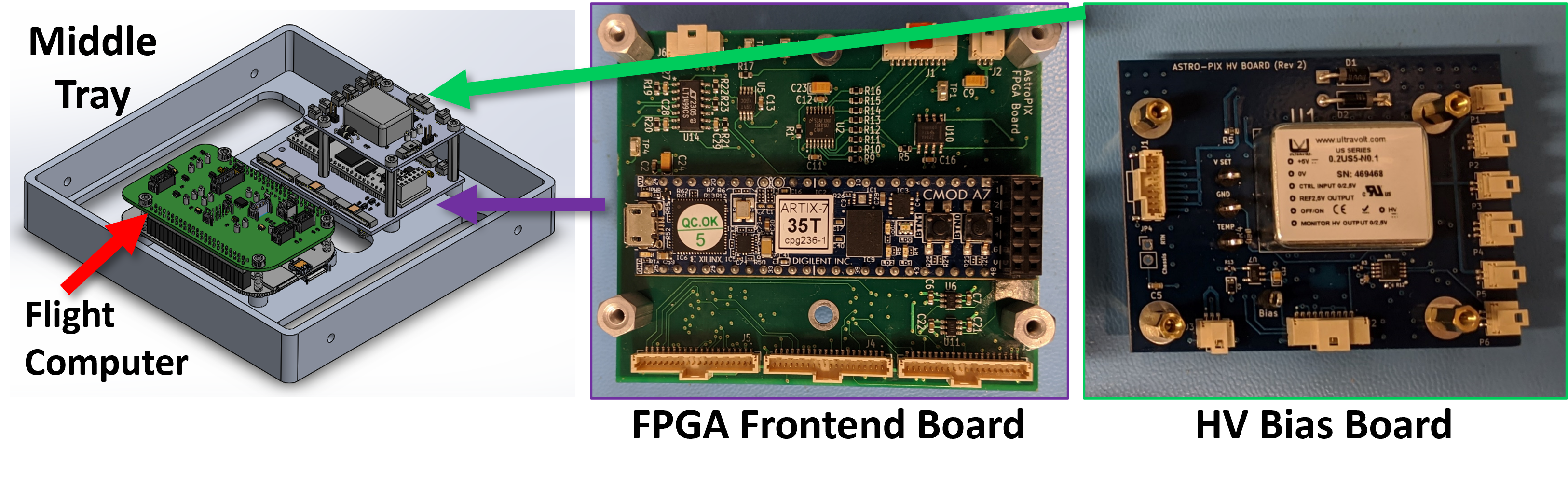}
    \caption{\textit{Left: }Middle layer of A-STEP payload containing frontend electronics and the flight computer. \textit{Middle: }FPGA Frontend Board that provides AstroPix quad chip command and data handling. \textit{Right: }A-STEP HV Bias board, providing a nominal -150\,V bias to AstroPix chips.}
    \label{fig:fig4}
\end{figure}

The frontend electronics of the A-STEP payload include the FPGA frontend board, and the HV supply board. Both of these boards are incorporated as a stack in the middle A-STEP payload tray as seen in Figure \ref{fig:fig4}.

The HV supply board receives secondary power from the PDU at 5.3\,V and regulates it down to  4.9\,V to drive the HV bias DC-DC converter. The HV bias voltage is set between 0\,V and $-$200\,V by a digital-to-analog converter (DAC) on the FPGA frontend board. Nominally, the AstroPix quad chips will each operate with a $-$150\,V bias during the sounding rocket flight. Independent HV supply and return lines are routed to each of the three stacked AstroPix carrier boards. The supplied bias voltage telemetry is measured through circuitry on the HV supply board and routed to an Analog-to-Digital (ADC) converter on the FPGA frontend board. This housekeeping data is read out alongside other temperature and current telemetry measurements.

The FPGA frontend board provides power, commanding, and data handling for the three AstroPix carrier boards. The FPGA frontend board receives power at 5.3\,V from the HV supply board and regulates it down to 4.5\,V. The board houses a Digilent CMOD A7 commercial-off-the-shelf (COTS) FPGA module with a Xilinx Artix 7 35T FPGA. The FPGA has been programmed with three parallel data interfaces routing to each of the stacked AstroPix carrier boards, containing both a quasi-SPI serial interface and a back-up shift register interface. Additional lines enable the AstroPix quad chips' operation, including lines for signal interrupt, hold, chip reset, and clocks for timestamping and the ToT measurement. On-board level shifters switch between the FPGA's 3.3\,V logic and AstroPix's 1.8\,V. Future iterations of the FPGA frontend board may opt for an FPGA solution with selectable voltage IO banks. 

The FPGA frontend board's ADC is used to digitize analog telemetry monitors throughout the payload for housekeeping data logging. The 8-channel, 12-bit ADC monitors the 4.5\,V supply current to each of the AstroPix carrier boards to detect shorts or unanticipated power draws. Additionally it monitors the HV bias provided to the AstroPix chips, and at least one thermal sensor on each AstroPix carrier board. 

\subsubsection{Flight Computer and Power}

The BeagleBone Black single-board computer has been chosen as the COTS flight computer for the A-STEP payload. The flight computer and accompanying flight software will be responsible for a couple crucial roles that begin autonomously after system boot. When the A-STEP instrument is powered on (prior to sounding rocket launch), the flight computer will:
\begin{enumerate}
    \item{Command the assignment of Chip IDs to each of the payload's 12 AstroPix sensors.}
    \item{Load locally saved, pre-selected configuration files to activate each AstroPix chip's bias block and disable known noisy pixels.}
    \item{Send a command to set the FPGA frontend board DAC to set the HV bias supply to $-$150\,V.}
    \item{Begin frequent (1\,Hz) collection of telemetry housekeeping data from the FGPA frontend board's ADC.}
    \item{Begin streaming data from the FPGA frontend board's event data buffer to the sounding rocket radio while creating a locally saved backup copy.}
\end{enumerate}

The flight computer connects with the FPGA frontend board for commanding and data transfer through an SPI interface. The flight computer will continue transmitting housekeeping and science data to the sounding rocket over a unidirectional EVTM through a User Datagram Protocol (UDP) link until the payload is powered off. In addition to the above interfaces, the flight computer is linked to the FPGA frontend board with ``cold" and ``warm" reset lines. The ``cold" reset line will prevent the FPGA from driving the flight computer's IOs at boot to prevent damage to the flight computer. The ``warm'' reset can be used by the flight computer to reset the FPGA free-running clock that is used to timestamp both housekeeping and science data. For more information on data types, see Section \ref{subsec:data}.

\begin{figure}[ht]
    \centering
    \includegraphics[width=0.8\textwidth]{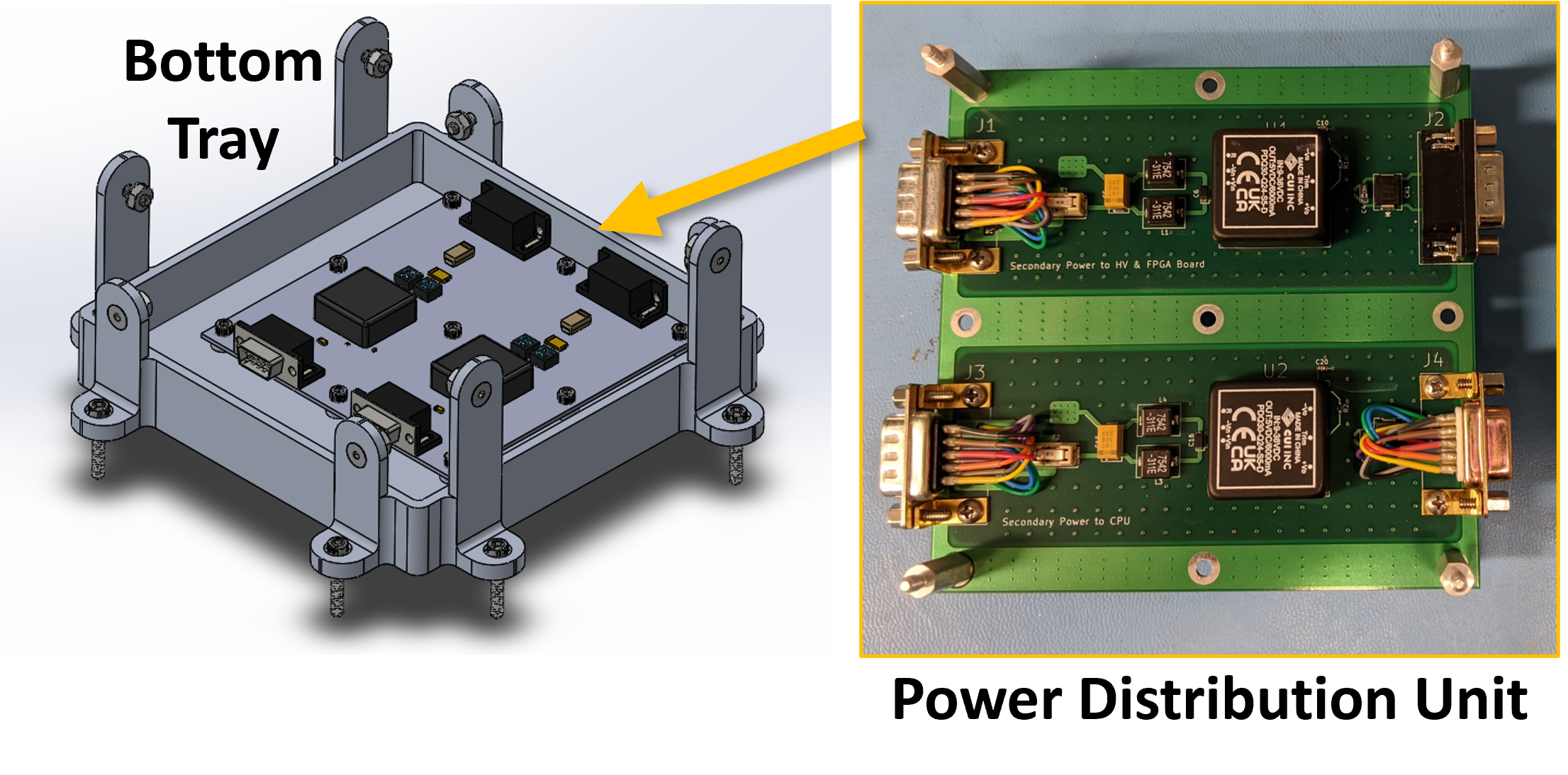}
    \caption{\textit{Left: }Bottom layer of A-STEP payload. \textit{Right: }Image of A-STEP Power Distribution Unit (PDU) board. Provides 5.0\,V for Beaglebone Flight Computer and a separate 5.3\,V supply for the remaining system.}
    \label{fig:fig5}
\end{figure}

The PDU receives 28\,V power from the sounding rocket bus and provides two output channels. The 5.3\,V output channel provides power and additional voltage headroom for the LDO power regulators on the HV bias board, the FPGA frontend board, and the AstroPix carrier boards. The second 5.0\,V power channel provides power directly to the BeagleBone black flight computer. Figure \ref{fig:fig5} shows the PDU separated from the remaining payload on the bottom tray for better thermal isolation. Aluminum blocks on the bottom tray will make direct contact with the PDU's DC-DC converters to improve heat dissipation.

\subsection{A-STEP Data Readout}
\label{subsec:data}

AstroPix chips have been designed with very large system integration in mind, which includes an efficient data transfer process. After a photon or charged particle event, the resulting charge is immediately digitized on-pixel and sent to a row and column buffer at the digital periphery. This quasi-SPI interface is described in more detail in Section \ref{subsec:astropix}. Each AstroPix carrier board has a single quasi-SPI interface and interrupt line that is routed to the FPGA frontend board.

The FPGA frontend board multiplexes the three AstroPix carrier board layer datastreams into a single buffered FIFO. While doing so, it prepends a header containing the AstroPix carrier board layer information and appends a timestamp from the FPGA's free-running clock. This timestamp, in conjunction with the 2 MHz timestamp clock provided to the AstroPix chips, allows sub-microsecond timing of photon events relative to the mission-elapsed time. This precise timestamping will allow A-STEP to meet is goal of linking events occurring between AstroPix carrier board layers to reconstruct particle tracks. For more information on the firmware that providing commanding and data-handling from the AstroPix chips and the AstroPix data packet format, consult the firmware documentation published by Richard Leys at KIT (\footnote{\url{https://astropix.github.io/astep-fw/}}).

The housekeeping data arrives from the FPGA frontend board in two formats. The first format is data from the 12-bit ADC, which includes the carrier board temperatures, 4.5\,V supply currents, and HV supply bias voltage values. Housekeeping data will additionally be provided by the FPGA frontend board, including FPGA internal temperatures, summed event rates, and other science data tracking metrics. On a fixed cadence (1 Hz) the flight software will collect housekeeping data and concatenate the various components into a single packet that is added to the EVTM data stream.

\section{Payload Integration \& Testing}
\label{sec:iandt}

Benchtop end-to-end payload testing will occur in Q3 2024, and include selected flight detectors, FPGA frontend board, HV bias board, and PDU. This configuration simulates a powered system similar to what will fly during the SubTEC-10 sounding rocket launch. Later in Q3, the flight computer and flight software will be incorporated into the benchtop end-to-end test and will be used to confirm autonomous instrument boot, configuration, and operation. At this time, A-STEP's mechanical enclosure will be fabricated, with full integration with the enclosure expected in Q4 2024.

Prior to delivery and integration into the launcher, A-STEP will be thoroughly tested to confirm that the AstroPix detectors and the other various payload electronics will survive sounding rocket launch and flight conditions. A-STEP will be operating in a sounding rocket compartment that includes a hatch that will open near apogee, exposing our payload to the space vacuum environment. Operation of the integrated A-STEP payload with vacuum chambers at GSFC prior to delivery will confirm that no unexpected outgassing, pressure differentials, or impacts to instrument performance may occur during this process.

\begin{figure}[ht]
    \centering
    \includegraphics[width=0.8\textwidth]{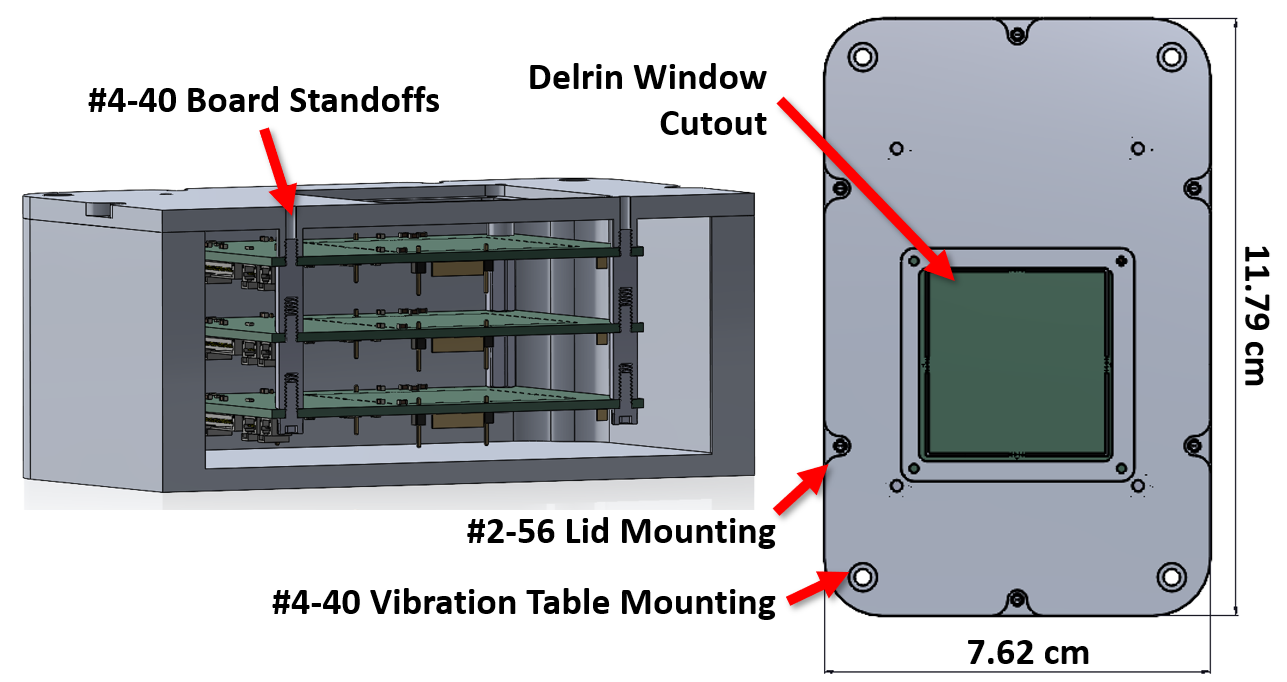}
    \caption{Mechanical fixture for vibration testing of detector carrier board stack.}
    \label{fig:fig6}
\end{figure}

The A-STEP payload will additionally undergo several vibration campaigns to ensure it will survive the sounding rocket launch conditions outlined in the NASA Sounding Rockets User Handbook \cite{SRHB}. In Q3, the AstroPix quad chip carrier boards will undergo vibration testing separately. This early testing will be a demonstration that the mounted AstroPix quad chip (with removed PCB insert) and associated wirebonds will survive launch. An AstroPix carrier board will first be tested in isolation. Following visual inspections and operational tests, vibration testing of the full three-board instrument stack will occur. See Figure \ref{fig:fig6} for the mechanical test enclosure designed for the AstroPix carrier board vibration tests. Pre- and post- vibration test visual inspections will confirm no change in AstroPix chip mounting, or wirebond contact. Pre- and post- vibration test operational checks will confirm that no unexpected shorts, changes in power, or degradation in AstroPix chip performance has occurred. After full payload assembly--including harnessing and the mechanical frame--is completed, another round of vibration testing will occur, with a final test campaign required after the payload is integrated into the sounding rocket.

\section{A-STEP Payload Simulations}
\label{sec:simulations}

Simulations were performed in order to evaluate A-STEP's ability to detect charged particle tracks through multiple detector layers during the sounding rocket flight. Particle interactions were simulated through Monte Carlo processes via Geant4 and MEGAlib toolkit's Cosima tool \cite{Zoglauer2006}. To run the simulation, detector models were generated in MEGAlib’s GeoMEGA tool utilizing the AstroPix quad chip geometries with active area of individual AstroPix chips (1.75$\times$1.75\,cm$^{2}$ and 70\,$\mu$m thick to conservatively match prior depletion depth measurements for {\tt AstroPixv3}) and AstroPix carrier board layer separation. Additionally, an aluminum box the approximate dimensions of the A-STEP mechanical enclosure was modelled.

\begin{figure}[ht]
    \centering
    \includegraphics[width=0.7\textwidth]{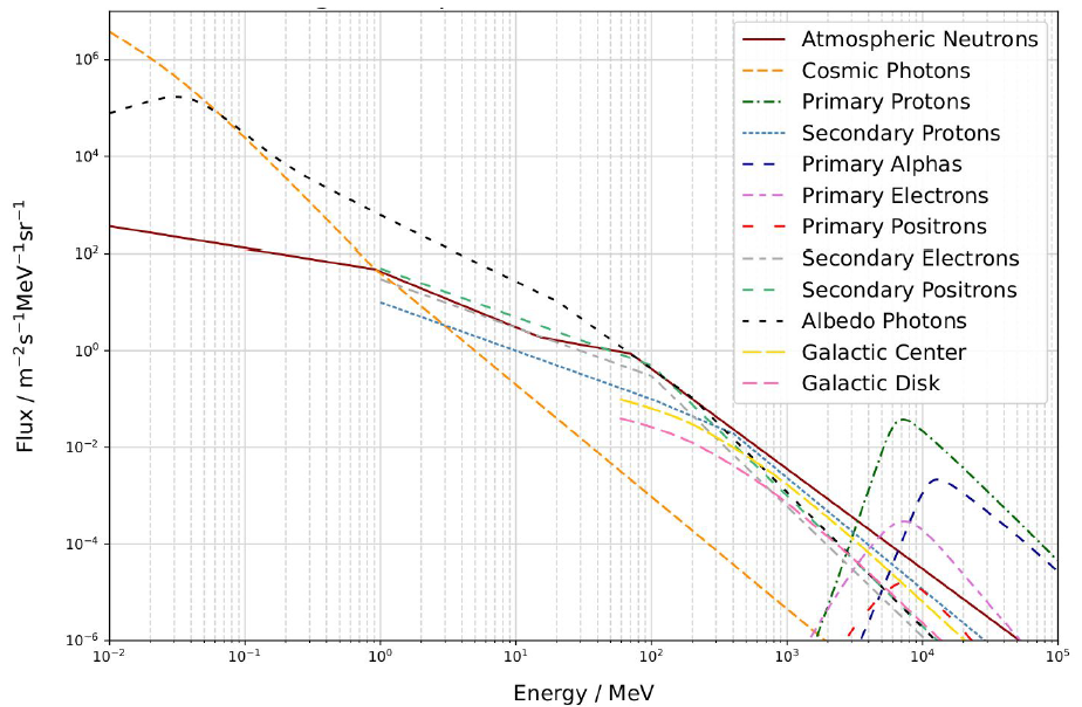}
    \caption{Background model spectra of a gamma-ray instrument at 300\,km and 30$^{\circ}$ inclination\cite{Cumani2019}.}
    \label{fig:fig7a}
\end{figure}

The particle environment simulated was based off a background model for low-Earth orbit gamma-ray satellites at 300\,km and an inclination of 30$^{\circ}$\cite{Cumani2019}. The particle background used in this simulation can be seen in Figure \ref{fig:fig7a}. The simulation considered an isotropic 4$\pi$ distribution of incident particle angles, as the exact orientation and role rate of the sounding rocket flight has not been defined. Each background particle type was simulated until 100,000 detector triggers were measured.

\begin{figure}[ht]
    \centering
    \hspace*{0cm}\includegraphics[width=0.7\textwidth]{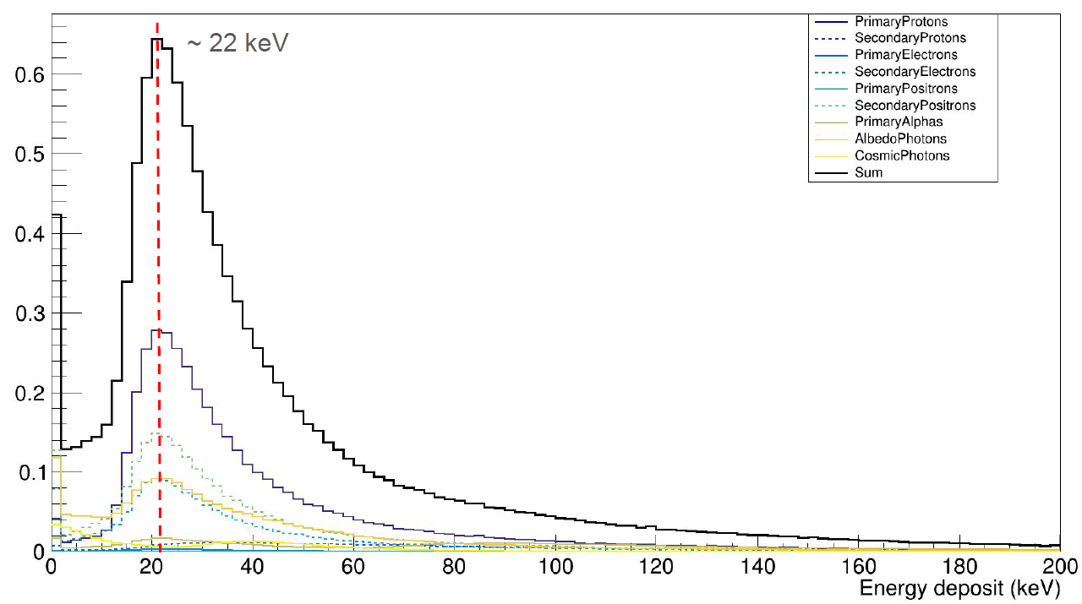}
    \caption{Distribution of energy deposited within the detector medium, normalized by detector trigger rates, peaking at 22\,keV (red dotted line).}
    \label{fig:fig7b}
\end{figure}

The cumulative deposited energy spectrum of the three AstroPix quad chip detector layers was computed by normalizing the deposited energy spectrum of each species by their trigger rate, and can be found in Figure \ref{fig:fig7b}. The cumulative energy distribution peaks at 22\,keV (red dotted line) which is low compared to the expected 25\,keV AstroPix pixel voltage comparator threshold setpoint. Comparative experiments to confirm the simulation results are planned. Potential methods to alleviate the threshold concerns have been identified, including operating the AstroPix chips in a selectable high-gain mode, or increasing the HV bias provided to improve pixel depletion.

\begin{figure}[ht]
    \hspace*{2cm}\includegraphics[width=0.8\textwidth]{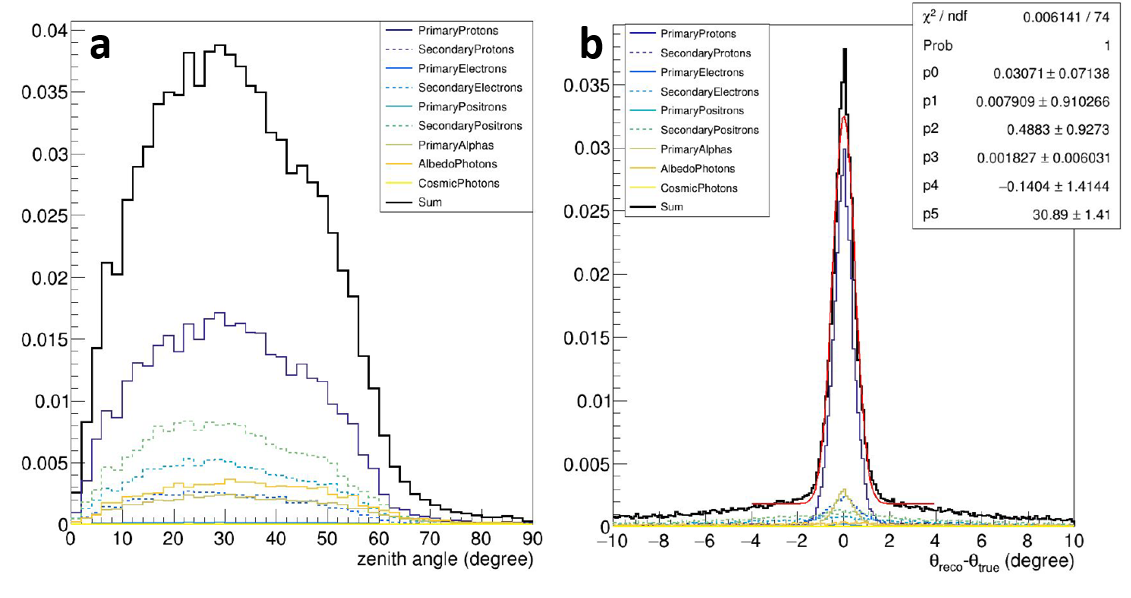}
    \caption{\textit{Left (a): }Total reconstructed zenith angle distribution of 3-layered events, normalized by trigger rates. \textit{Right (b): }Angular deviation distribution of all reconstructed 3-layered events, normalized by trigger rates. Fit with a double-Gaussian distribution.}
    \label{fig:fig8}
\end{figure}

The distribution of particles that generate three-layer events as a function of the particle’s incident angle with respect to zenith (0$^{\circ}$ is radially away from earth) can be seen in Figure \ref{fig:fig8}a, normalized by event trigger rates. The distribution peaks at approximately 30 degrees across all primary and secondary particle types, with primary protons and secondary electrons + positrons being the largest contributors to three-layer events. Figure \ref{fig:fig8}b shows the deviation between true particle trajectory and reconstructed particle trajectory for 3-layer events. The summed angular distribution was fit with a double gaussian, with fit parameters c$_1$, $\mu_1$, $\sigma_1$, c$_2$, $\mu_2$, $\sigma_2$ reported as p0--p5. 90\% of the reconstructed events fall within $\pm$ 6$^{\circ}$, with primary photons, primary alphas, and secondary protons yielding more accurate reconstructions than secondary particles. Additionally, particles with initial energies approximately $<$\,1 GeV are suffer from poor track reconstruction, as they often can cause multiple hits in a single tracker layer, resulting in multiple possible particle paths.

The results of the simulation efforts reveal that while treating the expected depletion region of 70\,$\mu$m as the total active thickness of the AstroPix detector geometry, we anticipate a total isotropic event trigger rate of 6.13\,Hz at an altitude of 300\,km before thresholding is applied. The primary contributors to this overall rate are primary protons (2.23\,Hz), albedo photons (1.10\,Hz), and secondary electrons (0.67\,Hz) + positrons (1.08\,Hz). Events that interact with all three AstroPix detector layers will occur at a 0.84\,Hz rate. Event rates are only marginally affected by adding or removing the aluminum rocket body from the overall geometry model. Primary protons are the largest contributor to the three-layer event rate and also the easiest to accurately reconstruct, followed by primary alphas and secondary protons. At the simulated total event rate, all data can be telemetered from the sounding rocket without any binning or prescaling requirements.

\section{A-STEP Launch Timeline}
\label{sec:timeline}

\begin{figure}[ht]
    \centering
    \includegraphics[width=0.8\textwidth]{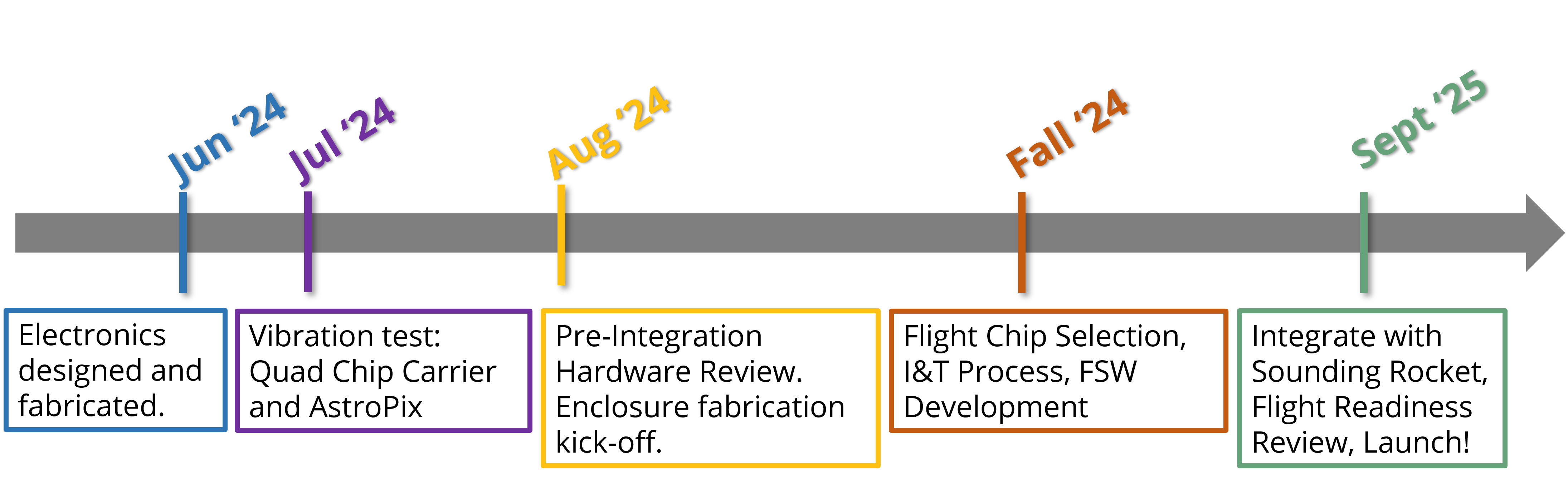}
    \caption{A-STEP Launch readiness timeline.}
    \label{fig:fig10}
\end{figure}

As of June of 2024, all electronic components of the A-STEP payload have been designed and fabricated. Final benchtop testing will occur before a pre-integration review of the system late in late summer 2024. A successful pre-integration review will trigger the fabrication of the full A-STEP mechanical enclosure. Q4 2024 through Q1 2025 will include the selection of the A-STEP flight AstroPix carrier boards (selected by fewest disabled/dead pixels and lowest achievable global threshold settings), integration of the A-STEP electrical system into the mechanical enclosure, and finalizing the development of the on-board flight software and associated ground-support graphical user interfaces and analysis tools. A full timeline of integration activities can be seen in Figure \ref{fig:fig10}.

A-STEP will launch on SubTEC-10 planned for September 2025. SubTEC-10 is a three-stage Terrier-Terrier-Improved Malemute that is predicted to reach an apogee of 260\,km high with an overall flight duration of 10 minutes. SubTEC-10 is anticipated to spend 380 seconds above 100\,km. A-STEP will be within a portion of the rocket that will be pumped down to a rough vacuum on the launchpad and that will open to space as it approaches apogee. All data will be telemetered down live during the launch, but data will additionally be saved locally on the flight computer in the event a water recovery of the rocket and payload is possible.

\section{Conclusion}
\label{sec:conclusion}

A-STEP is an important advancement in the multi-leveled, multi-program approach we are taking to certifying the technology readiness of the AstroPix HV-CMOS MAPS for use in future space-based observatories. We plan to take advantage of the low power and high tileability of AstroPix by adopting its design for a Compton-scatter and pair-production tracking telescope that will be incorporated into a next-generation medium-energy gamma ray observatory, such as AMEGO-X.

Through A-STEP, AstroPix sensors will be flown briefly in space with the goal of reconstructing charged particle tracks as a demonstration of the capabilities of the technology. A-STEP will launch as a technology demonstration payload on the SubTEC-10 sounding rocket mission from Wallops Island in September 2025.

\acknowledgments    
 
This work is funded by 18-APRA18-0084 and 20-RTF20-0003. DV’s research is funded through the NASA Postdoctoral Program through contract with ORAU. Beyond the authors of this publication, this work is supported by an international collaboration of scientists, engineers, technicians, and students who have all provided valuable assistance.

\bibliography{report} 

\begin{thebibliography}{10}

\bibitem{ASTRO2020}
{National Academies of Sciences, Engineering, and Medicine},  [{\em {Pathways to Discovery in Astronomy and Astrophysics for the 2020s}}{\nolinebreak\hspace{0.1em}]}, The National Academies Press, Washington, DC (2023).

\bibitem{Peric18}
{Peri{\'c}}, I. and {Berger}, N., ``{High Voltage Monolithic Active Pixel Sensors},'' {\em Nuclear Physics News}~{\bf 28},  25--27 (Jan. 2018).

\bibitem{Brewer2021}
{Brewer}, I., {Negro}, M., {Striebig}, N., {Kierans}, C., {Caputo}, R., {Leys}, R., {Peric}, I., {Fleischhack}, H., {Metcalfe}, J., and {Perkins}, J., ``{Developing the future of gamma-ray astrophysics with monolithic silicon pixels},'' {\em Nuclear Instruments and Methods in Physics Research A}~{\bf 1019},  165795 (Dec. 2021).

\bibitem{Steinhebel2022}
{Steinhebel}, A.~L., {Fleischhack}, H., {Striebig}, N., {Jadhav}, M., {Suda}, Y., {Luz}, R., {Kierans}, C., {Caputo}, R., {Tajima}, H., {Leys}, R., {Peric}, I., {Metcalfe}, J., and {Perkins}, J.~S., ``{AstroPix: novel monolithic active pixel silicon sensors for future gamma-ray telescopes},'' in [{\em Space Telescopes and Instrumentation 2022: Ultraviolet to Gamma Ray}{\nolinebreak\hspace{0.1em}]},  {den Herder}, J.-W.~A., {Nikzad}, S., and {Nakazawa}, K., eds., {\em Society of Photo-Optical Instrumentation Engineers (SPIE) Conference Series} {\bf 12181},  121816Y (Aug. 2022).

\bibitem{Striebig2024}
{Striebig}, N., {Leys}, R., {Peric}, I., {Caputo}, R., {Steinhebel}, A.~L., {Suda}, Y., {Fukazawa}, Y., {Jadhav}, M., {Violette}, D., {Kierans}, C., {Tajima}, H., {Metcalfe}, J., and {Perkins}, J.~S., ``{AstroPix4 {\textemdash} a novel HV-CMOS sensor developed for space based experiments},'' {\em Journal of Instrumentation}~{\bf 19},  C04010 (Apr. 2024).

\bibitem{Caputo2022}
{Caputo}, R., {Ajello}, M., {Kierans}, C.~A., {Perkins}, J.~S., {Racusin}, J.~L., {Baldini}, L., {Baring}, M.~G., {Bissaldi}, E., {Burns}, E., {Cannady}, N., {Charles}, E., {da Silva}, R. M.~C., {Fang}, K., {Fleischhack}, H., {Fryer}, C., {Fukazawa}, Y., {Grove}, J.~E., {Hartmann}, D., {Howell}, E.~J., {Jadhav}, M., {Karwin}, C.~M., {Kocevski}, D., {Kurahashi}, N., {Latronico}, L., {Lewis}, T.~R., {Leys}, R., {Lien}, A., {Marcotulli}, L., {Martinez-Castellanos}, I., {Mazziotta}, M.~N., {McEnery}, J., {Metcalfe}, J., {Murase}, K., {Negro}, M., {Parker}, L., {Phlips}, B., {Prescod-Weinstein}, C., {Razzaque}, S., {Shawhan}, P.~S., {Sheng}, Y., {Shutt}, T.~A., {Shy}, D., {Sleator}, C., {Steinhebel}, A.~L., {Striebig}, N., {Suda}, Y., {Tak}, D., {Tajima}, H., {Valverde}, J., {Venters}, T.~M., {Wadiasingh}, Z., {Woolf}, R.~S., {Wulf}, E.~A., {Zhang}, H., and {Zoglauer}, A., ``{All-sky Medium Energy Gamma-ray Observatory eXplorer mission concept},'' {\em Journal of Astronomical Telescopes, Instruments, and
  Systems}~{\bf 8},  044003 (Oct. 2022).

\bibitem{Apadula2022}
{Apadula}, N., {Armstrong}, W., {Brau}, J., {Breidenbach}, M., {Caputo}, R., {Carinii}, G., {Collu}, A., {Demarteau}, M., {Deptuch}, G., {Dragone}, A., {Giacomini}, G., {Grace}, C., {Graf}, N., {Greiner}, L., {Herbst}, R., {Haller}, G., {Jadhav}, M., {Joosten}, S., {Kenney}, C.~J., {Kierans}, C., {Kim}, J., {Markiewicz}, T., {Mei}, Y., {Metcalfe}, J., {Meziani}, Z.-E., {Nelson}, T.~K., {Peng}, C., {Pinaroli}, G., {Reimer}, P.~E., {Rota}, L., {Scott}, M., {Segal}, J., {Sichterman}, E., {Sinev}, N., {Steinhebel}, A., {Strom}, D., {Tricoli}, A., {Vernieri}, C., {Young}, C., and {Zurek}, M., ``{Monolithic Active Pixel Sensors on CMOS technologies},'' {\em arXiv e-prints} ,  arXiv:2203.07626 (Mar. 2022).

\bibitem{SRHB}
{Burth}, R., {Cathell}, P., {Edwards}, D., {Ghalib}, A., {Gsell}, J., {Hales}, H., {Haugh}, H., and {Tibbets}, B.,  [{\em {NASA Sounding Rockets User Handbook}}{\nolinebreak\hspace{0.1em}]}, Peraton Inc., NASA Wallops Flight Facility (May 2023).

\bibitem{Zoglauer2006}
{Zoglauer}, A., {Andritschke}, R., and {Schopper}, F., ``{MEGAlib The Medium Energy Gamma-ray Astronomy Library},'' {\em New Astronomy Review}~{\bf 50},  629--632 (Oct. 2006).

\bibitem{Cumani2019}
{Cumani}, P., {Hernanz}, M., {Kiener}, J., {Tatischeff}, V., and {Zoglauer}, A., ``{Background for a gamma-ray satellite on a low-Earth orbit},'' {\em Experimental Astronomy}~{\bf 47},  273--302 (June 2019).

\end{thebibliography}
\bibliographystyle{spiebib} 

\end{document}